\documentclass{nature1}

\usepackage{graphicx}
\usepackage{lineno}
\usepackage{physics}
\usepackage{xcolor}
\usepackage{float}
\usepackage{amsfonts}
\usepackage{amsmath}
\usepackage{amssymb}
\usepackage{bm}
\usepackage[section]{placeins}
\usepackage{fullpage}

\author
{Nicolas~Marquez~Peraca\textsuperscript{1,\textdagger},
Xinwei~Li\textsuperscript{2,*},
Jaime~M.~Moya\textsuperscript{1,3,\textdagger}, 
Kenji~Hayashida\textsuperscript{4,5,\textdagger},
Dasom~Kim\textsuperscript{3,5},
Xiaoxuan~Ma\textsuperscript{6}, 
Kelly~J.~Neubauer\textsuperscript{1}, 
Diego~Fallas~Padilla\textsuperscript{1}, 
Chien-Lung~Huang\textsuperscript{1,7}, 
Pengcheng~Dai\textsuperscript{1}, 
Andriy~H.~Nevidomskyy\textsuperscript{1}, 
Han~Pu\textsuperscript{1}, 
Emilia~Morosan\textsuperscript{1,8}, 
Shixun~Cao\textsuperscript{6,*}, 
Motoaki~Bamba\textsuperscript{9,10,11,*}, 
and~Junichiro~Kono\textsuperscript{1,5,12,*}
}

\begin{document}
\title{Quantum Simulation of an Extended Dicke Model \\with a Magnetic Solid}

\maketitle

\begin{affiliations}
\small
 \item Department of Physics and Astronomy, Rice University, Houston, TX 77005, USA
 \item Department of Physics, California Institute of Technology, Pasadena, CA 91125, USA
 \item Applied Physics Graduate Program, Smalley--Curl Institute, Rice University, Houston, TX 77005, USA
\item Division of Applied Physics, Graduate School of Engineering, Hokkaido University, Hokkaido, 060-8628, Japan
\item Department of Electrical and Computer Engineering, Rice University, Houston, TX 77005, USA
\item Department of Physics, International Center of Quantum and Molecular Structures, and Materials Genome Institute, Shanghai University, Shanghai, 200444, China
\item Department of Physics and Center for Quantum Frontiers of Research and Technology, National Cheng Kung University, Tainan 701, Taiwan
\item Department of Chemistry, Rice University, Houston, TX 77005, USA
\item Department of Physics I, Kyoto University, Kyoto 606-8502, Japan 
\item The Hakubi Center for Advanced Research, Kyoto University, Kyoto 606-8501, Japan
\item PRESTO, Japan Science and Technology Agency, Saitama 332-0012, Japan
\item Department of Materials Science and NanoEngineering, Rice University, Houston, TX 77005, USA
\end{affiliations}

\noindent\textsuperscript{\textdagger}{These authors contributed equally to this work}\newline
\noindent*Corresponding author. Email: kono@rice.edu (J.K.); bamba.motoaki.y13@kyoto-u.jp (M.B.); sxcao@shu.edu.cn (S.C.); xinweili@caltech.edu (X.L.)

\clearpage
%\linenumbers
\begin{abstract}
The Dicke model describes the cooperative interaction of an ensemble of two-level atoms with a single-mode photonic field and exhibits a quantum phase transition as a function of light--matter coupling strength.  Extending this model by incorporating short-range atom--atom interactions makes the problem intractable but is expected to produce new physical phenomena and phases. Here, we simulate such an extended Dicke model using a crystal of ErFeO$_3$, where the role of atoms (photons) is played by Er$^{3+}$ spins (Fe$^{3+}$ magnons). Through terahertz spectroscopy and magnetocaloric effect measurements as a function of temperature and magnetic field, we demonstrated the existence of a novel atomically ordered phase in addition to the superradiant and normal phases that are expected from the standard Dicke model.  Further, we elucidated the nature of the phase boundaries in the temperature--magnetic-field phase diagram, identifying both first-order and second-order phase transitions. These results lay the foundation for studying multiatomic quantum optics models using well-characterized many-body solid-state systems.
\end{abstract}

\newpage

\section*{Main text}

The Dicke model in quantum optics describes the cooperative, coherent coupling of an ensemble of two-level atoms with a single-mode light field\cite{Dicke1954}. Despite its simplicity, the model hosts a rich variety of phenomena that are significant in diverse contexts, such as cavity quantum electrodynamics\cite{Garraway2011}, condensed matter physics\cite{Cong2016}, and quantum information science\cite{FornDiaz2019,FriskKockum2019}. A prominent feature of the model is a second-order quantum phase transition (QPT), known as the superradiant phase transition (SRPT), which occurs when the light--matter coupling strength, $g$, exceeds a threshold\cite{HEPP1973360,Wang1973}. When the system enters the superradiant phase, atomic and photonic polarizations spontaneously emerge, producing a unique many-body ground state that enables studies of unusual light--matter entanglement\cite{Lambert2004}, two-mode squeezed states\cite{Artoni1991,Makihara2021,Hayashida2023}, and quantum chaos\cite{Emary2003}. 
	
Although the atomic ensemble in the original Dicke model was assumed to be noninteracting, %which is to some extent justifiable since a SRPT was faithfully simulated by a driven atom--cavity system with a Raman-type energy level scheme\cite{Baumann2010}.
it has been known from the early days that atom--atom interactions are important for explaining, for example, the dephasing and intensity correlation functions of fluorescent spectra\cite{Friedberg1974,Lawande1985}. Hence, there has long been interest in extending the Dicke model to include an atom--atom interaction (represented by strength $J$); see Fig.\,1. Such an extended Dicke model, or the $g$--$J$ model, should display an interplay of two types of interatomic interactions -- i.e., the photonic-field-mediated \emph{long-range} interaction, and the direct \emph{short-range} interaction. Intuitively, one can expect the ground state of the system to crucially depend on the ratio $g/J$, with a superradiant phase (an atomically ordered phase) favored for large (small) $g/J$. However, no analytical solutions can be obtained for the $g$--$J$ model, motivating one to simulate it using a well-characterized many-body quantum system.
    
Computational studies of the $g$--$J$ model under various approximations have revealed an array of new phenomena, such as a first-order QPT\cite{Lee2004,Chen2010,Yang2019,HerreraRomero2022,Zhao2017}, a shift of the SRPT boundary\cite{Nie2009,Chen2006}, amplification of the integrablity-to-chaos transition\cite{Wang2022}, modifications of matter--matter entanglement\cite{Nie2009,RoblesRobles2015}, and alteration of the nature of an excited-state QPT\cite{Rodriguez2018,HerreraRomero2022}. %Mean-field solutions to the $g$--$J$ model framed in complex networks have implications even for social sciences\cite{Bazhenov2021}, describing opinion formation within the communicating agents of a social group sharing a common information field.  
To examine these phenomena, several experimental platforms, including atomic Bose--Einstein condensates\cite{Chen2007,RodriguezLara2011}, superconducting qubits\cite{Zhang2014,Tian2010}, and quantum dots\cite{Lee2004}, have been proposed as quantum simulators, but successful simulations have not been achieved.
	
Here, we present a novel protocol of using a crystal of erbium orthoferrite (ErFeO$_3$), an antiferromagnetic (AFM) insulator, as a solid-state quantum simulator of the $g$--$J$ model. The magnetic properties of ErFeO$_3$ are governed by the moments carried by the Er$^{3+}$ and Fe$^{3+}$ spin subsystems and their interplay\cite{White1969}. A previous study has revealed Dicke cooperativity in the Er$^{3+}$--Fe$^{3+}$ interaction\cite{Li2018}, demonstrating the resemblance of the magnetic Hamiltonian of ErFeO$_3$ to the Dicke Hamiltonian. Namely, the paramagnetic Er$^{3+}$ ions (the magnons of ordered Fe$^{3+}$ spins) play the role of the atomic ensemble (light field), and the spin--magnon interaction is formally similar to the $g$-term in the Dicke model. What further strengthens this analogy is a magnetic phase transition of the crystal that exhibits all traits that would be expected for a Dicke SRPT. When the temperature ($T$) becomes lower than 4~K, the Er$^{3+}$ lattice develops C-type AFM order\cite{Zic2021} (with the ferromagnetic chains along $z$), and a zone-boundary Fe$^{3+}$ magnon mode condenses, displacing the staggered moments away from the $x$-$z$ plane\cite{Deng2015,Klochan1975}; this corresponds to the emergence of atomic and photon polarizations in the standard SRPT. In Bertaut's notation, the magnetic transition is of the $\Gamma_2\to\Gamma_{12}$ type (Fig.\,2a). Mean-field calculations using a realistic spin model captures the simultaneous order parameter (OP) onset of both the Er$^{3+}$ and Fe$^{3+}$ spin components, $\langle\Sigma_z^-\rangle$ and $\langle S_y\rangle$, respectively (Fig.\,2b), indicating that the $\Gamma_2\to\Gamma_{12}$ transition is a magnonic SRPT\cite{Bamba2022}, with the $\Gamma_2$ and $\Gamma_{12}$ phases corresponding to the normal (N) and superradiant (S) phases, respectively. 

One way to observe the OP onset is to monitor the quasi-antiferromagnetic (qAFM) magnon mode of Fe$^{3+}$ spins through terahertz (THz) time-domain spectroscopy, which has been utilized to reveal the configuration of Fe$^{3+}$ ions in rare-earth orthoferrites\cite{Li2022}. By performing THz transmission measurements on a $z$-cut ErFeO$_3$ crystal in the Faraday geometry, we obtained absorption coefficient ($\alpha$) spectra, derived from the imaginary part of the refractive index (Supplementary Section 2), as a function of $T$, as shown in Fig.\,2c. The observed bright absorption line is the qAFM mode, which has been thoroughly studied in previous studies\cite{Li2022}. It is the evolution of this mode in distinct phases of the $g$--$J$ model that is of interest throughout this study. A continuous OP-like onset, or a kink, is observed at the N\,$\to$\,S transition boundary ($<4$~K, blue dashed line). The frequency shift of the qAFM magnon mode in the S phase from that in the N phase is thus a sensitive reporter of the qAFM magnon condensate density, namely, the Fe$^{3+}$ OP of the S phase.

The $J$-term is inherently built into the magnetic Hamiltonian of ErFeO$_3$ since the Er$^{3+}$--Er$^{3+}$ exchange interaction, albeit being weak, is known to be present\cite{Kadomtseva1980}. Spectroscopic measurements have also revealed a fine frequency splitting within the Er$^{3+}$ electron paramagnetic resonance lines\cite{Li2018}, which is attributable to the Er$^{3+}$--Er$^{3+}$ exchange interaction. The presence of both the $g$- and $J$-terms sets the stage for ErFeO$_3$ to simulate the $g$--$J$ model. Nonetheless, although the $g$-term-driven S phase can find correspondence to the $\Gamma_{12}$ phase in ErFeO$_3$, the $g/J$ ratio set for the crystal stipulates that a pure atomic (A) phase, which is driven exclusively by the $J$-term, would not appear in equilibrium. For ErFeO$_3$, the A phase would be an Er$^{3+}$ ordered phase \emph{without involving any OP onset in the Fe$^{3+}$ subsystem}. Therefore, to achieve quantum simulation of the $g$--$J$ model, we must search for a way to invoke an explicit A phase through an S\,$\to$\,A transition.

Our theoretical consideration suggests that subjecting ErFeO$_3$ to a static magnetic field ($H$) along the $z$ axis can potentially induce an S\,$\to$\,A transition. This can be understood by writing the simplified magnetic Hamiltonian (Supplementary Section 1) in the second-quantized form as
%%% 
    \begin{equation}
		\begin{aligned}
\hat{\mathcal{H}}/\hbar=\omega_{\pi}\hat{a}_\pi^\dagger\hat{a}_\pi+\omega_\text{Er}\hat{\Sigma}_x^++\omega_z\hat{\Sigma}_z^++g\sqrt{\frac{2}{N_0}}i(\hat{a}_\pi^\dagger-\hat{a}_\pi)\hat{\Sigma}_z^- \\
		+J\frac{6}{N_0\hbar}[(\hat{\Sigma}_x^+)^2+(\hat{\Sigma}_z^+)^2-(\hat{\Sigma}_x^-)^2-(\hat{\Sigma}_z^-)^2],
		\label{totH}
		\end{aligned}
	\end{equation}
%%%
where a two-sublattice approximation is adopted for both Er$^{3+}$ and Fe$^{3+}$ for a total of $N_0$ unit cells. Here, $\omega_{\pi}$, $\hat{a}_\pi^\dagger$, and $\hat{a}_\pi$ are the energy, creation and annihilation operators for the Fe$^{3+}$ qAFM magnon mode, respectively; $\omega_\text{Er}$ is the frequency of Er$^{3+}$ spins as two-level systems at $H=0$; $\omega_z=|\mathfrak{g}_{ z }\mu_\text{B}\mu_0H|/\hbar$, where $\mathfrak{g}_{z}$ is the Land\`e $\textsl{g}$ factor, $\mu_\text{B}$ is the Bohr magneton, and $\mu_0$ is the vacuum permeability, is the $H$-induced Zeeman frequency of Er$^{3+}$; and $g$ and $J$ are the Er$^{3+}$--magnon and Er$^{3+}$--Er$^{3+}$ coupling strengths, leading to the $g$- and $J$-terms of the $g$--$J$ Hamiltonian, respectively.  
$\hat{\Sigma}_p=\sum_{i=1}^{2N_0}\hat{\sigma}_{i,p}/2$, where $\hat{\sigma}_p$ are Pauli matrices and $p\in\{x,y,z\}$, is the collective Er$^{3+}$ spin operator, with its superscript ``+" (``$-$") denoting the sum (difference) of the two sublattices. The way these operators appear in Eq.\,(\ref{totH}) is crucial for interpreting the ground-state energetics. Specifically, the $g$-term features a product of the Fe$^{3+}$ magnon field operator $i(\hat{a}_\pi^\dagger-\hat{a}_\pi)$ and the $\hat{\Sigma}_z^-$ component of Er$^{3+}$ spins, thereby favoring antiparallel alignment of Er$^{3+}$ sublattices and Fe$^{3+}$ magnon condensation in the S phase (the onsets of $\langle\hat{\Sigma}_z^-\rangle$ and $\langle S_y\rangle$ in Fig.\,2b), whereas the $J$-term couples Er$^{3+}$ antiferromagnetically; larger $\langle\hat{\Sigma}_x^-\rangle$ and $\langle\hat{\Sigma}_z^-\rangle$, where $\langle...\rangle$ denotes expectation values, are energetically more favorable.  

It is important to note that supplying the Zeeman term $\hat{\mathcal{H}}_\text{Zeeman}/\hbar=\omega_z\hat{\Sigma}_z^+$ provides quantum controllability. The term promotes $|\langle\hat{\Sigma}_z^+\rangle|$, the net moment of Er$^{3+}$ sublattices, through Zeeman coupling to $H\parallel z$. Due to the commutation relation
	\begin{equation}
	[\hat{\Sigma}_z^+, \hat{\Sigma}_z^-]=0\neq[\hat{\Sigma}_z^+, \hat{\Sigma}_x^-],
	\label{commutation}
	\end{equation}	
modification to $\langle\hat{\Sigma}_z^+\rangle$ would impact $\langle\hat{\Sigma}_x^-\rangle$ much more than $\langle\hat{\Sigma}_z^-\rangle$. This would tip the balance between the $g$-term and the $J$-term, since $\hat{\Sigma}_x^-$ appears only  in the $J$-term but not in the $g$-term.

As shown in Fig.\,3a, an S\,$\to$\,A transition is indeed recovered in the calculated mean-field phase diagram of the spin Hamiltonian (Supplementary Section 1) within the $T$-$H$ parameter space, for $T<2.8$~K, with a critical field ranging from 0.35~T to 0.5~T, depending on $T$. Increasing the field to above 1~T and elevating $T$ to above 4~K would both push the system across the thermodynamic phase boundary into the N phase. A triple point (at $2.8$~K and $0.5$~T, decorated by a yellow star) marks the location where the S, A, and N phases converge. Figure\,3b shows the calculated normalized spin components as the OPs of the magnetic phases, for a line cut along the $H$ axis at $T=0$~K, traversing sequentially the S\,$\to$\,A and the A\,$\to$\,N boundaries. 
%We identify that the Fe$^{3+}$ OP, represented by $\langle S_y\rangle$ (which is proportional to the magnon condensate density), is finite both in the S and A phases, but $\langle S_y\rangle\neq0$ in the S phase and $\langle S_y\rangle\approx0$ in the A phase. 
We identify that the Fe$^{3+}$ OP, represented by $\langle S_y\rangle$, is finite in the S phase but near-zero in the A phase.
The Er$^{3+}$ OP, on the other hand, is finite in both the S and A phases, but undergoes a switch from the $\langle\hat{\Sigma}_z^-\rangle\neq0$, $\langle\hat{\Sigma}_x^-\rangle\approx0$ type (S phase) to the $\langle\hat{\Sigma}_z^-\rangle\approx0$, $\langle\hat{\Sigma}_x^-\rangle\neq0$ type (A phase). Further, the OP evolution indicates that the S\,$\to$\,A boundary is an abrupt-type, first-order phase transition, while the A\,$\to$\,N boundary is a continuous-type, second-order phase transition.

Summarizing the mean-field calculation results, Fig.\,3c pictorially shows the predicted Fe$^{3+}$ and Er$^{3+}$ spin order in each phase. Starting from the N phase, the two sublattices of Fe$^{3+}$ are antiparallel along $z$ with zero $y$-component, while Er$^{3+}$ spins remain paramagnetic (no order). The A phase is characterized by Fe$^{3+}$ order that is identical to that of the N phase, but the Er$^{3+}$ subsystem develops canted AFM order where the sublattice moments are antiparallel along $x$ ($\langle\hat{\Sigma}_x^-\rangle\neq0$), with canting along $z$ ($\langle\hat{\Sigma}_z^+\rangle\neq0$). In the S phase, the Er$^{3+}$ order takes the $\langle\hat{\Sigma}_x^+\rangle\neq0$, $\langle\hat{\Sigma}_z^-\rangle\neq0$ configuration, and the staggered moment of the Fe$^{3+}$ sublattices undergoes a rotation about the $x$ axis, bringing its $y$-component to nonzero.

The S\,$\to$\,A transition can be considered as a spin-flop transition in terms of Er$^{3+}$ ions. One conventional way to characterize the transition is to monitor the magnetic susceptibility through which the existence of the AFM ordering of Er$^{3+}$ spins in the A phase has been previously observed\cite{Danshin1986}, although the configuration of Fe$^{3+}$ spins was left ambiguous. Our magnetization measurements showed clear S\,$\to$\,A and S\,$\to$\,N phase boundaries (squares in Fig. 4a and Supplementary Section 2). However, a strong and non-uniform demagnetizing effect that broadens the phase boundary\cite{Dudko1972} likely prevented us from clearly identifying the A\,$\to$\,N phase boundary. This is because the shape of our sample for magnetization measurements was a thin irregularly shaped disk cut from the sample used for THz measurements, rather than a sphere, which would have produced a uniform demagnetizing field inside the sample\cite{Danshin1986}. Nonetheless, a disk-shaped sample with a large lateral size was necessary for performing THz transmission measurements.

To demonstrate the A\,$\to$\,N phase boundary, i.e., the breakdown of the AFM order of Er$^{3+}$ spins, we performed magnetocaloric effect (MCE) experiments which are sensitive to the magnetic entropy landscape of a material. Namely, the Gr\"uneisen ratio\cite{zhu2003universally}

\begin{equation}
\Gamma_H~=~-\frac{(\partial {S}/\partial {H})_T}{C_H}~=~\frac{1}{T}\frac{\partial {T}}{\partial{H}}\bigg|_{S},
\end{equation}
% $Γ_H= -(∂M/∂T)_H/C_H = -(∂S/∂H)_T/(T(∂S/∂T)_H )=1/T  (∂T/∂H)_S

\noindent measures the slope of isentropes in the $T -H$ plane\cite{garst2005sign}. Since the heat capacity $C_H$ is always a positive quantity, the sign of $(\frac{\partial {T}}{\partial{H}})_S$ is always opposite to $(\frac{\partial {S}}{\partial {H}})_T$. Furthermore, sharp changes in entropy $S$ due to phase transitions will appear as step functions in $(\frac{\partial {T}}{\partial{H}})_S$\cite{garst2005sign}, or peaks if $(\frac{\partial {T}}{\partial{H}})$ is measured in a quasi-adiabatic environment\cite{Jaime2002}. Thus, by measuring the differential change in sample temperature with respect to the magnetic field,  $(\frac{\partial {T}}{\partial{H}})_S$, the $T-H$ magnetic phase diagram can be measured. We note, that the demagnetization factor can have a small effect on the MCE measurements\cite{romero2014influence}, namely that temperature and field shifts can occur, but the qualitative features should be present. 

The $T$-$H$ phase diagram of ErFeO$_3$, and the obtained results are summarized in Fig.\,4a. %; the result shows {\color{red}overall?} agreement with the phase diagram {\color{red}computed by the mean-field theory}. 
%MCE measurements capture the differential change in sample temperature with respect to the magnetic field, $\mu^{-1}_0(\partial T/\partial H)_S$.
We configured a MCE measurement in a Physical Property Measurement System in the quasi-adiabatic condition\cite{moya2022field}, and took raw data traces of sample temperature variation versus magnetic field at a ramping rate of 5\,$\times$\,10$^{-3}$~T/s with $dH > 0$ (Supplementary Section 2); the sensitivity of temperature variation of our instrument reached 5\,$\times$\,10$^{-4}$~K. To identify $H$-induced phase transitions, the first-order derivative  $(\frac{\partial {T}}{\partial{H}})_S$ was approximated as $dT/d(\mu_0 H)$ %traces were computed 
(Supplementary Section 2), whose local extremes correspond to the transition boundaries\cite{Jaime2002}. The traces clearly exhibit two maxima for $T<2.8$~K, corresponding to the S$\to$A (red dashed line in Fig.\,4a) and A\,$\to$\,N (blue dashed line) boundaries, and one maximum for $2.8$~K~$<T<4$~K, corresponding to the S\,$\to$\,N (blue dashed line) boundary. These results are qualitatively consistent with the $T-H$ phase diagram reported previously\cite{Danshin1986}, where quantitative shifts likely come from demagnetization effects.
	
Once we experimentally investigated the evolution of the atomic ensemble (or the Er$^{3+}$ spins) in the extended Dicke Hamiltonian, we turned to elucidate the photonic counterpart (or the Fe$^{3+}$ spins). The ambiguity of the configuration of Fe$^{3+}$ spins and the nature of the transition boundaries require us to monitor the qAFM magnon mode of Fe$^{3+}$ spins in THz magnetospectroscopy experiments. Unlike the static measurements, responses from different domains in the A phase can be distinguished in the frequency domain, illuminating the nature of the phase transition. The measurements were performed within the same $T$-$H$ parameter space as that of the MCE experiments. Figures~4b-d and Figs.\,4e-g show the $H$-dependence of $\alpha$ spectra at select $T$ values and the $T$-dependence of $\alpha$ spectra at select $\mu_0 H$ values, respectively. We found that the bright absorption lines can be assigned to either the quasi-ferromagnetic (qFM) mode or the qAFM mode\cite{Li2022}, the latter of which can be an OP for the Fe$^{3+}$ spins. 
	
In the $H$-dependent color map at $1.4$~K (Fig.\,4b), three lines are observed. The lowest frequency line, which does not pick up intensity until 0.8~T, is the qFM mode, while the other two are both qAFM magnons, albeit belonging to distinct phases. The middle (upper) line, which is located at 0.8~THz at 0~T (1~THz at 0.5~T), is the qAFM mode of the S (A \& N) phase. The S\,$\to$\,A transition can be identified to occur at $0.5$~T (red dashed line), where the upper line emerges. The qAFM magnons belonging to the S and A phases coexist within $0.5$~T~$<\mu_0 H<1$~T, consistent with the prediction that the S\,$\to$\,A transition is of first order and is thus inhomogeneous, until the middle line vanishes at $>$1~T (blue dashed line) owing to entrance into the N phase. The $3.2$~K map (Fig.\,4c) shows a different behavior; the qAFM magnon (0.88~THz at 0~T) of the S phase continuously shifts to connect with that of the N phase in frequency, forming an OP-like onset for $\mu_0 H<0.7$~T (blue dashed line), signaling a second-order N\,$\to$\,S transition boundary. Such a frequency shift is absent in the $4.4$~K map (Fig.\,4d) since the N phase persists throughout the whole $H$ range.
	
$T$-dependent color maps at constant $H$ further corroborate our assignments of the phase transitions. Again from the $0$~T map (Fig.\,4e), a continuous OP-like onset of the qAFM mode shift is observed ($<4$~K, blue dashed line). This echoes Fig.\,4c in showing the continuous nature of the N\,$\to$\,S transition, and establishes that the frequency shift of the qAFM magnon in the S phase from that in the N phase can be the Fe$^{3+}$ OP of the S phase. Intriguingly, this OP is demonstrated to be zero in the A phase. We read this fact from the $0.75$~T map (Fig.\,4f), for which an N\,$\to$\,A transition is expected upon lowering $T$. Although a residual mode pertaining to the S phase exists (as mentioned earlier when discussing Fig.\,4b), the qAFM mode (unlabeled line) frequency does not undergo any noticeable OP-like anomaly across the N\,$\to$\,A transition; it is as featureless as the qAFM mode within the $1.25$~T map (Fig.\,4g), for which the N phase persists throughout the whole $T$ range. This unambiguously demonstrates that the spin order in the A phase only involves Er$^{3+}$ ordering but not any Fe$^{3+}$ OP, consistent with our expectation depicted in Fig.\,3. Finally, phase boundaries determined by the THz experiments are overlaid (as solid circles) on top of the MCE phase diagram in Fig.\,4a, showing overall agreement.

A potential impact of this analogy is the possibility of being applied to other members of the rare-earth orthoferrite family or orthochromite compounds. For example, spin-reorientation phase transitions\cite{Belov1976,balbashov1995submillimeter} ($\Gamma_4\to\Gamma_2$) in RFeO$_3$ (R=Yb, Er, and Tb) would mimic the SRPT. In YbFeO$_3$, where the Yb$^{3+}$--Yb$^{3+}$ interaction ($J$) is negligible, it would be a potential playground for studying the standard Dicke model ($g$ model). At the boundaries of the phase transition of YbFeO$_3$, the qFM mode of Fe$^{3+}$ shows a kink, and a transition inside the ground doublet of Yb$^{3+}$ ions shows a softening\cite{Danshin1986Yb}. This simultaneous kink and softening is one of the hallmarks of a magnonic SRPT\cite{Kim2024}. It was also suggested that TbFeO$_3$ can be regarded as the magnetic phase transition of the Jahn-Teller type\cite{Zvezdin1976,Belov1979} that would resemble a magnonic SRPT. In ErFeO$_3$, where a crystal field transition ($\sim$1.5\,THz) is responsible for the spin-reorientation transition ($T$ = 87\,K), the crystal field levels would play the role of an ensemble of two-level atoms in the Dicke model. To prove that Dicke physics is at work, however, one must show Dicke cooperativity, i.e., the coupling strength $g$ must exhibit cooperative enhancement $g\propto\sqrt{N}$, where $N$ is the number of two-level atoms. In addition, mapping their spin Hamiltonians into the Dicke models is required to establish this analogy. No attempts have been made to develop an analogy between the spin-reorientation transition and the Dicke superradiant phase transition.

The advantages of using the low-temperature phase transition, as opposed to the $\sim$80\,K spin-reorientation phase transition, of ErFeO$_3$ in simulating the extended Dicke model can be summarized as follows. First, the low-temperature phase transition allows us to simulate the first-order phase transition into the A phase, which is the main point of this work and does not exist in the spin-reorientation transition at 87\,K. Second, since we deal with the lowest two energy levels (Kramers doublet) of Er$^{3+}$ ions in the low-temperature phase transition, theoretical analysis is directly relevant to the Dicke model, compared to the multiple crystal-field energy levels involved in the 87\,K phase transition. Third, and most importantly, at high temperatures, thermally populated magnons are not negligible. Such thermal magnons will prevent studies of the vacuum magnons responsible for the Dicke superradiant phase transition, which occurs in thermal equilibrium without any external driving. For example, one consequence of the superradiant phase transition induced by vacuum bosonic fields is a two-mode perfect squeezed vacuum at the critical point\cite{Bamba2022}. A finite number of thermally excited magnons will mask such interesting quantum phenomena.

In summary, through THz magnetospectroscopy and magnetocaloric effect experiments, we studied a crystal of ErFeO$_3$ to simulate the $g$--$J$ model, which is an extended Dicke model that includes not only the bosonic-field-mediated long-range interatomic interactions but also direct short-range interactomic interactions. In addition to the superradiant and normal phases expected from the standard Dicke model, we identified a new phase, an atomic phase, which is driven by the short-range $J$-term in the Hamiltonian.  Further, we elucidated the nature of the various phase boundaries, distinguishing between first-order and second-order transitions.  These results demonstrated the potential of ErFeO$_3$ as a simulator of quantum optics Hamiltonians.  More specifically, in the context of Dicke physics, this condensed matter platform may lead to the possibilities of assisting quantum chaos\cite{Wang2022} and modifying matter--matter entanglement\cite{Nie2009,RoblesRobles2015} with tunability given through an external magnetic field.  Bridging the gap between quantum optics and many-body correlated physics, our results will find broader application in the design of hybrid quantum systems with superior controllability, such as the Dicke-Ising machine\cite{Rohn2020,Zhang2014} and the Dicke-Lipkin-Meshkov-Glick model\cite{RoblesRobles2015,HerreraRomero2022}. Furthermore, the ability to transition between the superradiant and atomic phases via a nonthermal knob provides opportunities to study unconventional quantum criticality\cite{Xu2019} and chaos-assisted thermalization\cite{Altland2012}.

\section*{Methods}

Methods, including statements of data availability and any associated accession codes and references, are available in Supplementary Information.

%\nolinenumbers
\clearpage
\section*{References}
\bibliographystyle{sn-nature}
\bibliography{EFO_LTPT}

\section*{Acknowledgements}
We thank Shiming Lei, Andrey Baydin, Takuma Makihara, Fuyang Tay, and Timothy Noe for useful discussions. J.K.\ acknowledges support from the U.S.\ Army Research Office (through Award No.\ W911NF2110157), the W.\ M.\ Keck Foundation (through Award No.\ 995764), the Gordon and Betty Moore Foundation (through Grant No.\ 11520), and the Robert A.\ Welch Foundation (through Grant No.\ C-1509). X.L.\ acknowledges support from the Caltech Postdoctoral Prize Fellowship and the IQIM. S.C.\ is grateful for financial support from the National Natural Science Foundation of China (NSFC, No.\ 12074242), and the Science and Technology Commission of Shanghai Municipality (No.\ 21JC1402600). M.B.\ acknowledges support from the JST PRESTO program (Grant JPMJPR1767). J.M.M.\ was supported by the National Science Foundation Graduate Research Fellowship under Grant DGE 1842494. A.H.N.\ was supported by the Robert A.\ Welch Foundation (through Grant No.\ C-1818) and the US National Science Foundation (through Grant No.\ DMR-1917511). P.D. was supported by U.S. DOE BES DE-SC0012311.

\section*{Author contributions} N.M.P.\ performed THz measurements and analyzed all THz data under the guidance and supervision of X.L.\ and J.K.\  J.M.M., C-L.H., and E.M.\ performed magnetization and MCE measurements and data analysis and discussed the results with N.M.P, X.L., D.K., and J.K. K.H.\ and M.B.\ calculated spin resonance frequencies, spin configurations, oscillator strengths, and mean-field phase diagrams.  M.B.\ supervised K.H.\ in the theoretical modeling. X.M.\ grew, cut, and characterized the high-quality ErFeO$_3$ single crystals used in the experiments under the guidance of S.C. K.J.N.\ and P.D.\ performed additional Laue diffraction measurements. D.F.P., H.P., and A.H.N. contributed to the theoretical analysis. N.M.P., X.L., J.M.M., K.H., D.K., M.B., and J.K. wrote the manuscript. All authors discussed the results and commented on the manuscript.

\section*{Competing interests} 
The authors declare no competing financial interests.

\section*{Additional information} Supplementary information is available. Correspondence and requests for materials should be addressed to J.K.

\section*{Data availability and code availability}
Data and code that support the findings of this study are available from the corresponding author upon reasonable request.

	\begin{figure}[p]
		\centering
		\includegraphics[width=0.5\linewidth]{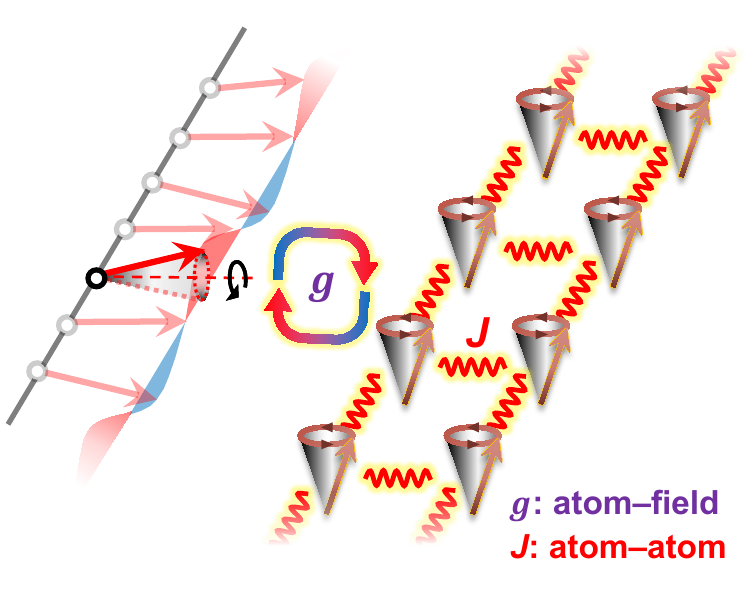}
		\caption{\small \textbf{The extended Dicke model, or the $g$--$J$ model, where an ensemble of interacting two-level atoms collectively couples with a bosonic field.} The cooperative boson--atom interaction, with strength $g$, mediates long-range atom--atom interactions, whereas the direct atom--atom interaction, with strength $J$, is short-ranged.
		}
		\label{Fig1}
	\end{figure}

    \begin{figure}[p]
		\centering
		\includegraphics[width=0.5\linewidth]{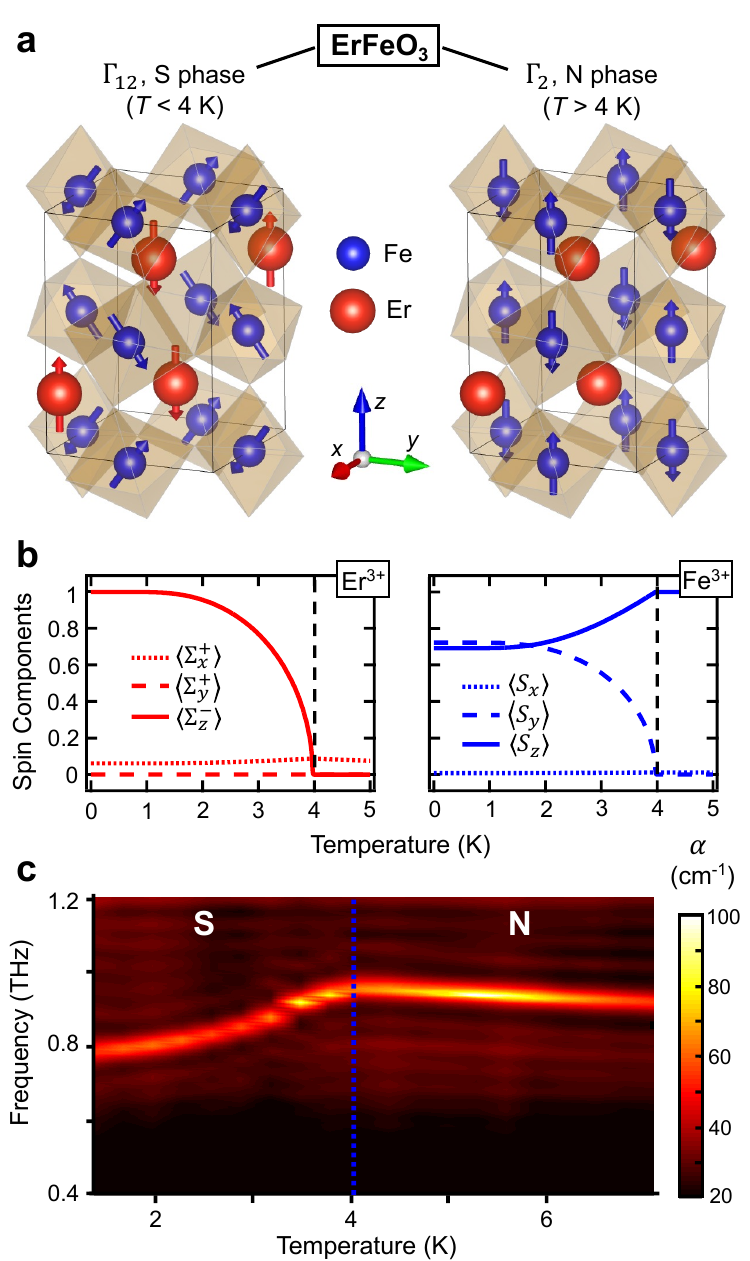}
		\caption{\small \textbf{The $\Gamma_2\to\Gamma_{12}$ transition in ErFeO$_3$ as a magnonic analog of the SRPT.} \textbf{a},~Lattice structure and spin configurations within the $\Gamma_{12}$ and $\Gamma_2$ phases. Brown polyhedra represent octahedrally coordinated FeO$_6$ cages. \textbf{b},~Temperature dependence of the Er$^{3+}$ and Fe$^{3+}$ spin components (normalized) across the phase transition at 0\,T. $\hat{\Sigma}_p$, where $p\in\{x,y,z\}$, is the collective Er$^{3+}$ spin operator, with its superscript ``+" (``$-$") denoting the sum (difference) of the two sublattices. $\hat{S}_p$ are the components of Fe$^{3+}$ spins. \textbf{c},~Temperature dependence of THz absorption spectra taken at zero magnetic field. The bright line, corresponding to the qAFM magnon mode of Fe$^{3+}$ spins, shows a kink at 4\,K, which is the superradiant-normal phase boundary at zero magnetic field.
        }
		\label{Fig2}
	\end{figure}

	\begin{figure}[p]
		\centering
		\includegraphics[width=0.5\linewidth]{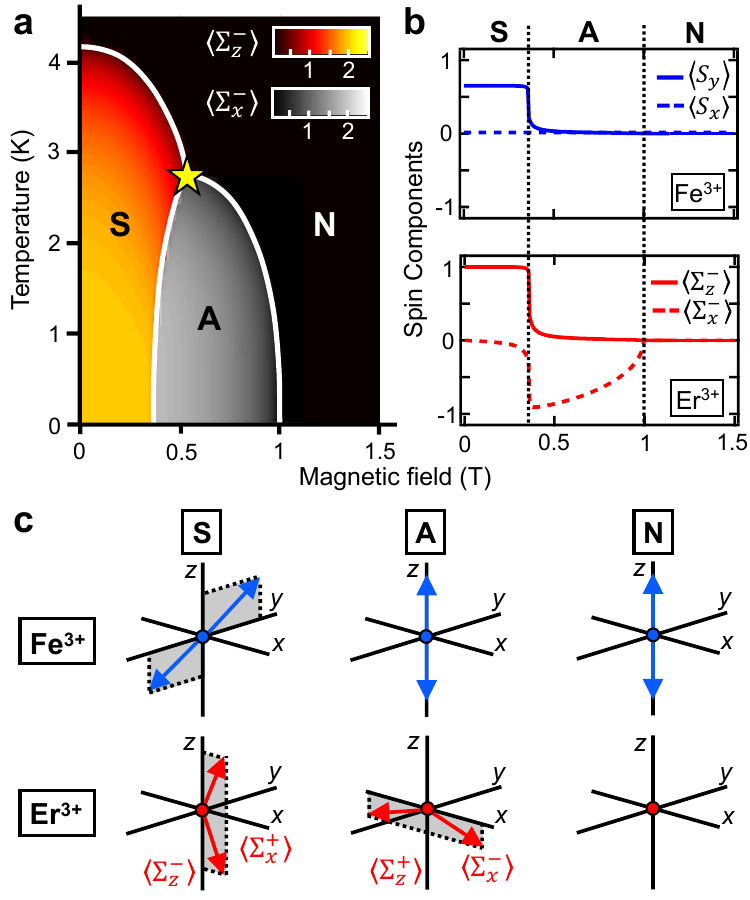}
		\caption{\small \textbf{Mean-field solution for the spin Hamiltonian of ErFeO$_3$ in $H \parallel z$.} \textbf{a},~Theoretical $T$-$H$ phase diagram mapped by Er$^{3+}$ spin components. \textbf{b},~$H$-dependent evolution of the Er$^{3+}$ and Fe$^{3+}$ spin components (normalized) at $T=0$~K. \textbf{c},~Schematic diagrams of the spin configuration in each phase. 
		}
		\label{Fig3}
	\end{figure} 

	\begin{figure}[p]
	\centering
	\includegraphics[width=1\linewidth]{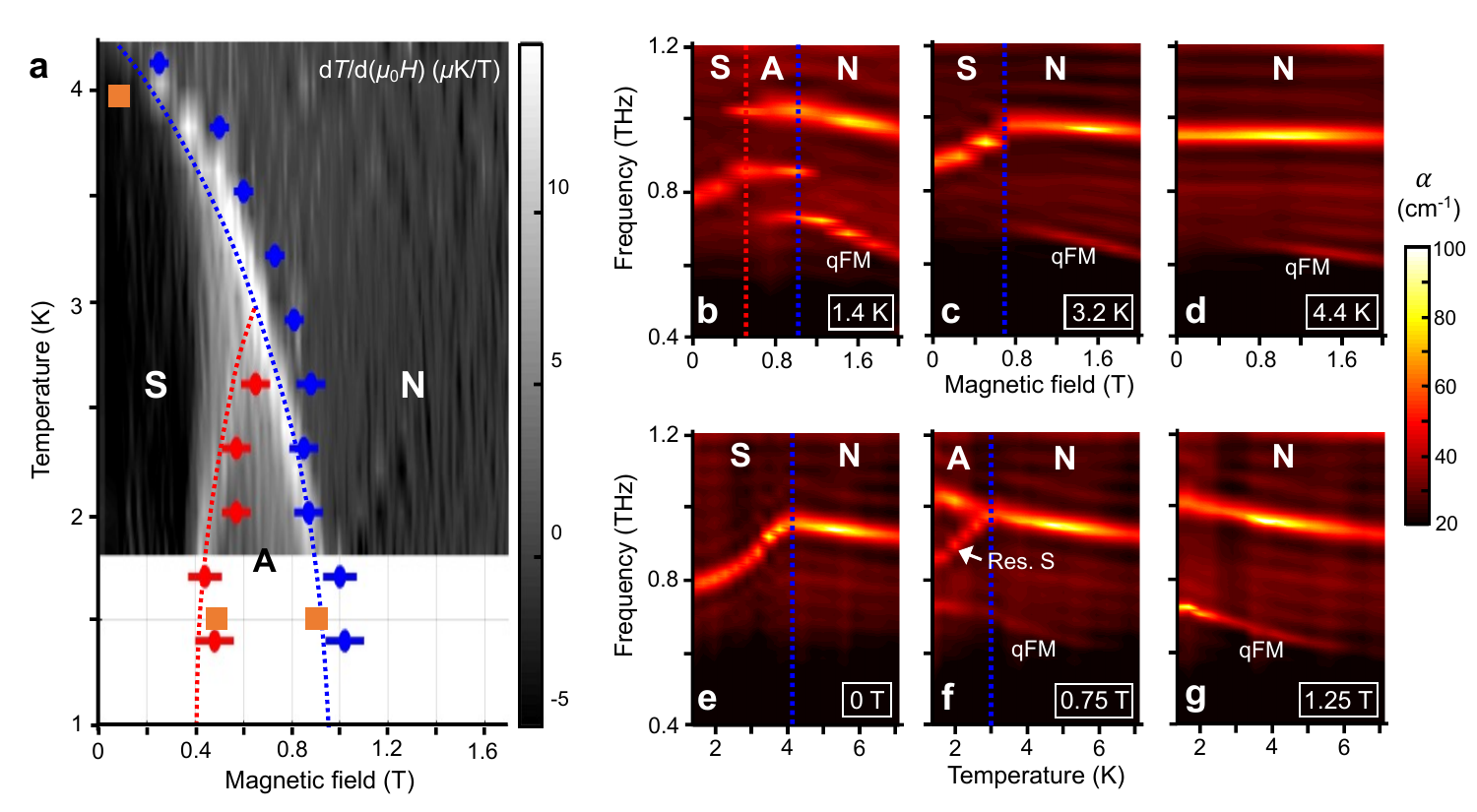}
	\caption{\small \textbf{Mapping out the $T$-$H$ phase diagram of ErFeO$_3$ in $H\parallel z$.} \textbf{a},~Phase boundaries determined by THz measurements (solid circles) and magnetization measurements (squares) overlaid on the $dT/d(\mu_0H)$ color map determined from MCE measurements. Red (blue) dashed line denotes a first- (second-) order phase boundary. \textbf{b}-\textbf{d},~THz absorption spectra mapped vs $\mu_0 H$ for select $T$ values. \textbf{e}-\textbf{g},~THz absorption spectra mapped vs $T$ for select $\mu_0 H$ values. Red and blue dashed lines mark the same boundaries as those in \textbf{a}. All features except for those labeled ``qFM" are qAFM magnon modes of Fe$^{3+}$ spins.
	}
	\label{Fig4}
 % (b)~MCE data traces at select temperatures (1.8~K to 4.2~K with 0.4~K interval) showing $T-T_0$ ($T_0$ being the initial temperature) vs $\mu_0H$ (left) and $dT/d(\mu_0H)$ vs $\mu_0 H$ (right); curves are offset. Red and blue arrows mark the boundaries outlined by red and blue dashed lines in (a), respectively. 
 
\end{figure}

\end{document}